\documentclass[conference,draftcls,onecolumn]{IEEEtran}

\usepackage{amsmath,amsfonts,amssymb,bbm,amsthm} 
\usepackage{multirow}
\usepackage{extarrows}
\usepackage{algorithm}
\usepackage{algorithmic}
\usepackage{enumerate}
\usepackage{threeparttable}
\usepackage{color}
\ifCLASSINFOpdf
\usepackage[pdftex]{graphicx}
\else
\fi
\usepackage{float}
\usepackage{cite}
\usepackage{epsfig}
\usepackage{mathrsfs}
\usepackage{array}
\usepackage{subfig}
\usepackage{verbatim}
\usepackage{setspace}
\usepackage{nicefrac}
\usepackage{arydshln}
\usepackage{changepage}
\usepackage[dvipsnames]{xcolor}
\usepackage{url}
\usepackage{fancyhdr}                                
\usepackage{lastpage}                                           
\usepackage{layout}

\DeclareGraphicsExtensions{.eps,.pdf,.jpg,.png}

\DeclareMathOperator*{\argmin}{argmin}

\newtheorem{theorem}{Theorem}
\newtheorem{lemma}{Lemma}

\newtheorem{definition}{Definition}

\allowdisplaybreaks


\newcommand\bovermat[2]{%
	\makebox[0pt][l]{$\smash{\overbrace{\phantom{%
					\begin{matrix}#2\end{matrix}}}^{\text{#1}}}$}#2}

\begin{document}

\title{Fundamental Limits of Approximate Gradient Coding}

%\author{Sinong Wang, Jiashang Liu and Ness Shroff
%\thanks{Sinong Wang, Jiashang Liu and Ness Shroff were with the Department
%of Electrical and Computer Engineering, The Ohio State University, Columbus,
%OH, 43210 USA e-mail: \{wang.7691, liu.3992 shroff.11\}@osu.edu}}
\author{\IEEEauthorblockN{Sinong Wang$^{*}$, Jiashang Liu$^{*}$ and Ness Shroff$^{*\dag}$}
	\IEEEauthorblockA{$^{*}$Department of Electrical and Computer Engineering\\
		$^{\dag}$Department of Computer Science and Engineering\\
		The Ohio State University\\
		\{wang.7691, liu.3992 shroff.11\}@osu.edu}}

\maketitle

\begin{abstract}
It has been established that when the gradient coding problem is distributed among $n$ servers, the computation load (number of stored data partitions) of each worker is at least $s+1$ in order to resists $s$ stragglers~\cite{tandon2017gradient}. This scheme incurs a large overhead when the number of stragglers $s$ is large. In this paper, we focus on a new framework called \emph{approximate gradient coding} to mitigate stragglers in distributed learning. We show that, to exactly recover the gradient with high probability, the computation load is lower bounded by $O(\log(n)/\log(n/s))$. We also propose a code that exactly matches such lower bound.  We identify a fundamental three-fold tradeoff for any approximate gradient coding scheme $d\geq O(\log(1/\epsilon)/\log(n/s))$, where $d$ is the computation load, $\epsilon$ is the error of gradient. We give an explicit code construction based on random edge removal process that achieves the derived tradeoff. We implement our schemes and demonstrate the advantage of the approaches over the current fastest gradient coding strategies.	
\end{abstract}

\section{Introduction}

Large-scale machine learning has shown great promise for solving many practical applications~\cite{le2011optimization}. Such applications require massive training datasets and model parameters, and force practitioners to adopt distributed computing frameworks such as Hadoop~\cite{dean2008mapreduce} and Spark~\cite{zaharia2010spark} to increase the learning speed. However, the speedup gain is far from ideal due to the latency incurred in waiting for a few slow or faulty processors, called ``straggler'' to complete their tasks~\cite{dean2012large}. For example, it was observed in~\cite{yadwadkar2016multi} that a straggler may run $8\times$ slower than the average worker performance on Amazon EC2. 
To alleviate the straggler issue, current frameworks such as Hadoop deploy various straggler detection techniques and usually replicate the straggling tasks on other available nodes.
%Recently, the coding theoretic techniques provide a more effective way to deal with ``straggler'' in the distributed learning. In this paper, we focus on a coding theoretic techniques for mitigating stragglers in distributed learning. 

Recently, \emph{gradient coding} techniques have been proposed to provide an effective way to deal with straggler for distributed learning applications~\cite{tandon2017gradient}. The system being considered has $n$ workers, in which the training data is partitioned into $n$ parts. Each worker stores multiple parts of datasets, computes a partial gradient over each of its assigned partitions, and returns the linear combination of these partial gradients to the master node. By creating and exploiting coding redundancy in local computation, the master node can reconstruct the full gradient even if part of results are collected, and therefore alleviate the impact of straggling workers.

The key performance metric used in gradient coding scheme is the \emph{computation load} $d(s)$, which refers to the number of data partitions that are sent to each node, and characterizes the amount of redundant computations to resist $s$ stragglers. Given the number of workers $n$ and number of stragglers $s$, the work of~\cite{tandon2017gradient} establishes a fundamental bound $d(s)\geq s+1$, and constructs a random code that exactly matches this lower bound. Two subsequent works~\cite{dutta2016short,raviv2017gradient} provide a deterministic construction of the gradient coding scheme. These results imply that, to resist one or two stragglers, the best gradient coding scheme will double or even triple the computation load in each worker, which leads to a large transmission and processing overhead for data-intensive applications.

In practical distributed learning applications, we only need to \emph{approximately} reconstruct the gradients. For example, the gradient descent algorithm is internally robust to the noise of gradient evaluation, and the algorithm still converges when the error of each step is bounded~\cite{bottou2010large}. In other scenarios, adding the noise to the gradient evaluation may even improve the generalization performance of the trained model~\cite{neelakantan2015adding}. These facts motivate the idea of \emph{approximate gradient coding} technique. More specifically, suppose that $s$ of the $n$ workers are stragglers, the approximate gradient coding allows the master node to reconstruct the full gradient with a multiplicative error $\epsilon$ from $n-s$ received results. The computation load in this case is a function $d(s,\epsilon)$ of both number of stragglers $s$ and error $\epsilon$. By introducing the error term, one may expect to further reduce the computation load. Given this formulation, we are interested in the following key questions: 

%In this paper, we develop an idea of approximate coding technique, we call \emph{approximate gradient coding}, for mitigating stragglers with a lower computation load. We show that this coding scheme recovers a noisy gradient with high probability, instead of exactly recovering the full gradient for any $s$ stragglers. The basic motivation is: (i) gradient descent method consists of multiple iterations; if one can operate most steps with significant less computation load, both the job completion time and computation overhead can be drastically decreased; (ii) the gradient descent algorithm is internally robust to the noise of gradient evaluation~\cite{bottou2010large}, and the algorithm still converges when the error of each step is bounded. In this paper, we are interested in the following key questions: given that $s$ of the $n$ workers are stragglers, and a constraint on the error being less than $\epsilon$ for each gradient evaluation.

\emph{What is the minimum computation load for the approximate gradient coding problem? Can we find an optimal scheme that achieves this lower bound?}

There have been two computing schemes proposed earlier for this problem. The first one, introduced in~\cite{raviv2017gradient}, utilizes the expander graph, particularly Ramanujan graphs to provide an approximate construction that achieves a computation load $O(ns/(n-s)\epsilon)$ given error $\epsilon$. However, expander graphs, especially Ramanujan graphs, are expensive to compute in practice, especially for large number of workers. Hence, an alternative computing scheme was recently proposed in~\cite{charles2017approximate}, referred to as \emph{Bernoulli Gradient Code} (BGC). This coding scheme incurs a computation load of $O(\log(n))$ and an error of $O(n/(n-s)\log(n))$ with high probability. 

\subsection{Main Contribution}
In this paper, we show that, the optimum computation load can be far less than what the above two schemes achieve. More specifically, we first show that, if we need to exactly ($\epsilon=0$) recover the full gradients with high probability, the minimum computation load satisfies
\begin{equation}
d(s,0)\geq O\left(\frac{\log(n)}{\log(n/s)}\right).
\end{equation}
We also design a coding scheme, referred to as $d$\emph{-fractional repetition code} (FRC) that achieves the optimum computation load. This result implies that, if we allow the decoding process to fail with a vanishing probability, the computation load in each worker can be significantly reduced from $s+1$ to $O(\log(n)/\log(n/s))$. For example, when $n=100$ and $s=10$, each worker in the original gradient coding strategy requires storing $11\times$ data partitions, while approximate scheme requires only $2\times$ data partitions. 

Furthermore,  we identify the following three-fold fundamental tradeoff among the computation load $d(s,\epsilon)$, recovery error $\epsilon$ and number of stragglers $s$ in order to approximately recover the full gradients with high probability. The tradeoff reads
\begin{equation*}
d(s,\epsilon)\geq O\left(\frac{\log(1/\epsilon)}{\log(n/s)}\right).
\end{equation*}
This result provides a quantitative characterization that the noise of gradient plays a logarithmic reduction role, i.e., from $O(\log(n))$ to the $O(\log(n))-O(\log(\epsilon n))$ in the desired computation load. For example, when the error of gradient is $O(1/\log(n))$, the existing BGC scheme in~\cite{charles2017approximate} provides a computation load of $O(\log(n))$, instead, the information-theoretical lower bound is $O(\log(\log(n)))$.
We further give an explicit code construction, referred to as \emph{batch raptor code} (BRC), based on random edge removal process that achieves this fundamental tradeoff. The comparison of our proposed schemes and existing gradient coding schemes are listed in TABLE~\ref{tab:comparison}.

We finally implement and benchmark the proposed gradient coding schemes at Ohio Supercomputer center~\cite{OhioSupercomputerCenter1987} and empirically demonstrate its performance gain compared with existing strategies. 

\begin{table}[t]
	\small
	\vskip -0.1in
	\centering
	\caption{Comparison of Existing Gradient Coding Schemes}
	\vskip 0.05in
	\label{tab:comparison}
	\begin{threeparttable}
		\centering
		\begin{tabular}{|c|c|c|c|}
			\hline
			Scheme & Computation Load  & Error of gradient \\
			& load & gradient\\
			\hline
			cyclic MDS~\cite{tandon2017gradient} & $s+1$ & $0$ \\
			\hline
			expander graph code~\cite{raviv2017gradient} & $O\left(\frac{ns}{(n-s)\epsilon}\right)$&$\epsilon$ \\
			\hline
			BGC\tnote{1}~\cite{charles2017approximate}  & $O(\log(n))$  & $O\left(\frac{n}{(n-s)\log(n)}\right)$  \\
			\hline
			FRC\tnote{1} & $O\left(\frac{\log(n)}{\log(n/s)}\right)$ & $0$\\
			\hline
			BRC\tnote{1} &  $O\left(\frac{\log(1/\epsilon)}{\log(n/s)}\right)$ & $\epsilon$ \\
			\hline
		\end{tabular}
		\begin{tablenotes}
			\scriptsize
			\item[1] result holds with high probability, i.e., 1-o(1).
		\end{tablenotes}
	\end{threeparttable}
	\vskip -0.25in
\end{table} 

\subsection{Related Literature}

The works of Lee et al.~\cite{lee2017speeding} initiated the study of using coding technique such as MDS code for mitigating stragglers in the distributed linear transformation problem and the regression problem. Subsequently, one line of studies was centered on designing the coding scheme in distributed linear transformation problem. Dutta et al.~\cite{dutta2016short} constructed a deterministic coding scheme in the product of a matrix and a long vector. Lee et al.~\cite{lee2017high} designed a type of efficient 2-dimensional MDS code for the high dimensional matrix multiplication problem. Yu et al.~\cite{yu2017polynomial} proposed the optimal coding scheme, named as polynomial code, in the matrix multiplication problem. Wang et al.~\cite{wang2018,wang2019} further initialized the study of computation load in the distributed transformation problem and design several efficient coding schemes with low density generator matrix.

The second line of researches focus on constructing the coding schemes in the distributed algorithm in machine learning application. The work of~\cite{karakus2017straggler} first addressed the straggler mitigation in linear regression problem by data encoding. Our results are closely related to designing the code for general distributed gradient descent or the problem of computing sum of functions. The initial study by~\cite{tandon2017gradient} presented an optimal trade-off between the computation load and straggler tolerance for any loss functions. Two subsequent works in~\cite{charles2017approximate,raviv2017gradient} considered the approximate gradient evaluation and proposed the BGC scheme with less computation load compared to the scheme in~\cite{tandon2017gradient}. Maity et al.~\cite{maity2018robust} applied the existing LDPC to a linear regression model with sparse recovery. Ye et al.~\cite{ye18} further introduced the communication complexity in such problem and constructed an efficient code for reducing both straggler effect and communication overhead. None of the aforementioned works characterizes the fundamental limits of the approximate gradient coding problem. In the sequel, we will systematically investigate this problem.

\section{Preliminaries}
\subsection{Problem Formulation}
\begin{figure}[htb]
	\vskip -0.1in
	\begin{center}
		\centerline{\includegraphics[width=3.6in]{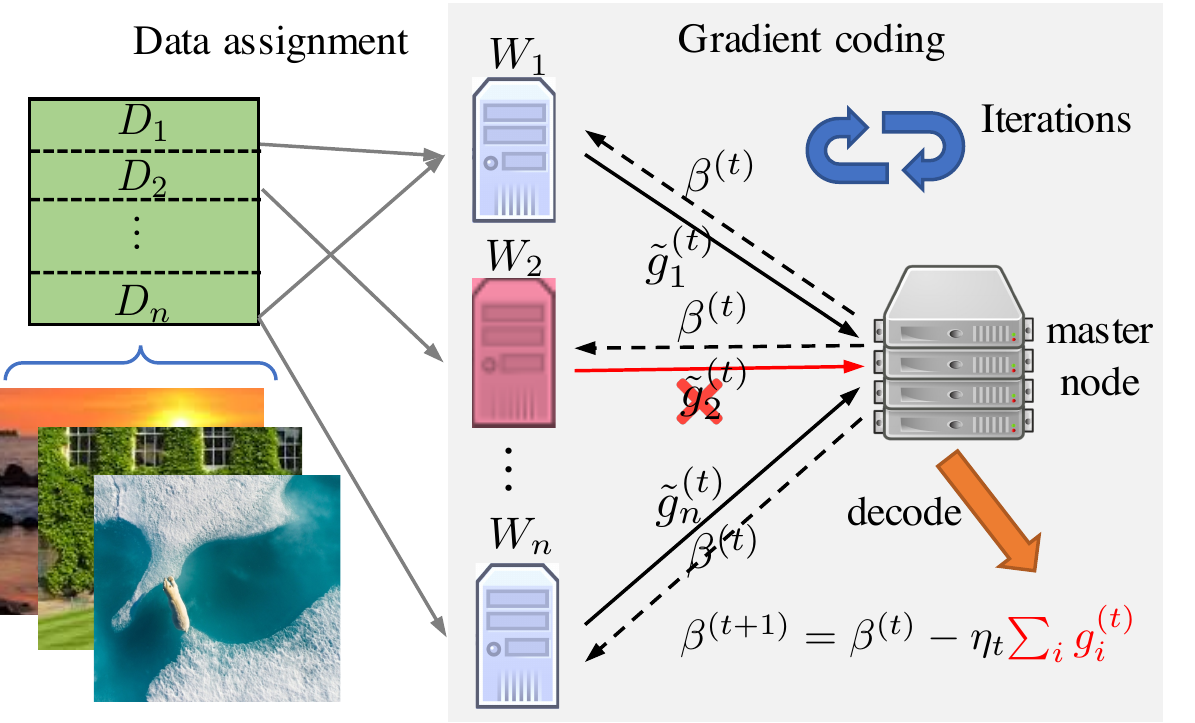}}
		\vskip -0.1in
		\caption{Gradient coding framework.}
		\label{fig:system}
	\end{center}
	\vskip -0.4in
\end{figure}

The data set is denoted by $D=\{(x_i,y_i)\}_{i=1}^N$ with input feature $x_i\in\mathbb{R}^p$ and label $y_i\in\mathbb{R}$. Most machine learning tasks aim to solve the following optimization problem:
\begin{equation}
\beta^*=\argmin\limits_{\beta\in\mathbb{R}^p}\sum\limits_{i=1}^N L(x_i,y_i;\beta)+\lambda R(\beta),
\end{equation}
where $L(\cdot)$ is a task-specific loss function, and $R(\cdot)$ is a regularization function. This problem is usually solved by gradient-based approaches. More specifically, the parameters $\beta$ are updated according to the iteration $\beta^{(t+1)}=h_R(\beta^{(t)},g^{(t)})$, where $h_R(\cdot)$ is the proximal mapping of gradient-based iteration, and $g^{(t)}$ is the gradient of the loss function at the current parameter $\beta^{(t)}$, defined as
\begin{equation}
g^{(t)}=\sum\limits_{i=1}^N \nabla L(x_i,y_i;\beta^{(t)}).
\end{equation}
In practice, the number of data samples $N$ is quite large, i.e., $N\geq 10^9$, the evaluation of the gradient $g^{(t)}$ will become a bottleneck of the above optimization process and should be distributed over multiple workers. Suppose that there are $n$ workers $W_1, W_2, \ldots, W_n$, and the original dataset is partitioned into $n$ subsets of equal size $\{D_1,D_2,\ldots,D_n\}$. In the traditional distributed gradient descent, each worker $i$ stores the dataset $D_i$. During iteration $t$, the master node first broadcasts the current classifier $\beta^{(t)}$ to each worker. Then each worker $i$ computes a partial gradient $g_i^{(t)}$ over data block $D_i$, and returns it to the master node. The master node collects all the partial gradients to obtain a gradient evaluation $g^{(t)}=\sum\nolimits_{i=1}^n g_i^{(t)}$ and updates the classifier correspondingly. In the gradient coding framework, as illustrated in Figure.~\ref{fig:system}, each worker $i$ stores multiple data blocks and computes a linear combination of partial gradients, then the master node receives a subset of results and decodes the full gradient $g^{(t)}$. 

More formally, the gradient coding framework can be represented by a coding matrix $A\in\mathbb{R}^{n\times n}$, where the $i$th worker computes\footnote{Here we omit iteration count $t$.} $\tilde{g}_i=\sum\nolimits_{j=1}^nA_{ij}g_j$. Let $g\in\mathbb{R}^{n\times p}$ ($\tilde{g}\in\mathbb{R}^{n\times p}$) be a matrix with each row being the (coded) partial gradient
\begin{equation*}
\tilde{g}=[\tilde{g}_1;\tilde{g}_2;\ldots;\tilde{g}_n] \text{ and } g=[g_1;g_2;\ldots;g_n].
\end{equation*}
Therefore, we can represent $\tilde{g}=Ag$. Suppose that there exist $s$ stragglers and the master node receives $n-s$ results indexed by set $S$. Then the received partial gradients can be represented by $\tilde{g}_S=A_Sg$, where $A_S\in\mathbb{R}^{(n-s)\times n}$ is a row submatrix of $A$ containing rows indexed by $S$. During the decoding process, the master node solves the following problem,
\begin{equation}\label{eq:decode_ls}
u^*=\argmin\nolimits_{u\in\mathbb{R}^{n-s}}\|A_S^Tu-1_n\|^2,
\end{equation}
and recovers the full gradient by $u^*\tilde{g}$, where $1_n\in\mathbb{R}^{n}$ denotes the all one vector. 
\begin{definition}\label{def:recovery_error}
	\emph{(Recovery error)} Given a submatrix $A_S\in\mathbb{R}^{(n-s)\times n}$, the corresponding recovery error is defined as
	\begin{equation}
	\emph{err}(A_S)=\min\limits_{u\in\mathbb{R}^{n-s}}\|A_S^Tu-1_n\|^2,
	\end{equation}
\end{definition}
Instead of directly measuring the error of recovered gradient, i.e., $\min_u\|uA_Sg-1_ng\|$, this metric quantifies how close $1_n$ is to being in the span of the columns of $A_S$.  It is also worth noting that the overall recovery error is small relative to the magnitude of the gradient, since the minimum decoding error satisfies $\min_u\|uA_Sg-1_ng\|\leq\|g\|\cdot\min_u\|uA_S-1_n\|$.
\begin{definition}\label{def:computation_load}
	\emph{(Computation load)} The computation load of a gradient coding scheme $A$ is defined as $\kappa(A)=\max_{1\leq i\leq n}\|A_i\|_0$, where $\|A_i\|_0$ is the number of nonzero coefficients of the $i$th row $A_i$.
\end{definition}
The existing work~\cite{tandon2017gradient} shows that the minimum computation load is at least $s+1$ when we require decoding the full gradient exactly, i.e., err$(A_S)=0$, among all $S\subseteq [n], |S|=n-s$. The \emph{approximate gradient coding} relaxes the worst-case scenario to a more realistic setting, the ``average and approximate'' scenario. Formally, we have the following systematic definition of the approximate gradient codes.
\begin{definition}
	\emph{($\epsilon$-approximate gradient code)} Given number of $s$ stragglers in $n$ workers, the set of $\epsilon$-approximate gradient code is defined as 
	\begin{equation}
	\mathcal{G}_{\epsilon}=\{A\in\mathbb{R}^{n\times n}|\mathbb{P}[\emph{err}(A_S)>\epsilon n]=o(1)\},
	\end{equation}
	where  $A_S\in\mathbb{R}^{(n-s)\times n}$ is a randomly chosen row submatrix of $A$.
\end{definition}
The above definition of gradient code is general and includes most existing works on approximate gradient coding. For example, let $\delta=s/n$, the existing scheme based on 
Ramanujan graphs is a $\epsilon$-approximate gradient code that achieves computation load of $O(s/(1-\delta)\epsilon)$;
the existing BGC~\cite{charles2017approximate} can be regarded as a $O(1/(1-\delta)\log(n))$-approximate gradient code that achieves computation load of $O(\log(n)/(1-\delta))$.

\subsection{Main Results}

Our first main result provides the minimum computation load and corresponding optimal code when we want to exactly decode the full gradient with high probability.
\begin{theorem}\label{thm:main_1}
	Suppose that out of $n$ workers, $s = \delta n$ are stragglers. 
	The minimum computation load of any gradient codes in $\mathcal{G}_0$ satisfies
	\begin{equation}
	\kappa_0^*(A)\triangleq\min\limits_{A\in \mathcal{G}_0} \kappa(A) \geq O\left(\max\left\{1,\frac{\log(n)}{\log(1/\delta)}\right\}\right).
	\end{equation}
	And we construct a gradient code, we call $d-$fractional repetition code $A^{\text{FRC}}\in\mathcal{G}_0$, such that
	\begin{equation}
	\lim\limits_{n\rightarrow\infty}\kappa(A^{\text{FRC}})/\kappa_0^*(A)=1.
	\end{equation}
\end{theorem}
The following main result provides a more general outer bound when we allow the recovered gradient to contain some error.
\begin{theorem}\label{thm:tradeoff_error_load}
	Suppose that out of $n$ workers, $s = \delta n$ are stragglers. If $0<\epsilon<O(1/\log^2(n))$, the minimum computation load of any gradient codes in $\mathcal{G}_{\epsilon}$ satisfies
	\begin{equation*}
	\kappa_{\epsilon}^*(A)\triangleq\min\limits_{A\in \mathcal{G}_{\epsilon}} \kappa(A) \geq O\left(\max\left\{1,\frac{\log(1/\epsilon)}{\log(1/\delta)}\right\}\right).
	\end{equation*}
	And we construct a gradient code, named as \emph{batch raptor code} $A^{\text{BRC}}\in\mathcal{G}_c$, such that
	\begin{equation}
	\lim\limits_{n\rightarrow\infty}\kappa(A^{\text{BRC}})/\kappa_{\epsilon}^*(A)=1.
	\end{equation}
\end{theorem}
Theorem~\ref{thm:tradeoff_error_load} provides a fundamental tradeoff among the gradient noise, the straggler tolerance and the computation load. And the gradient noise $\epsilon$ provides a factor of logarithmic reduction, i.e., $-\log(\epsilon n)$ of the computation load. 

\textbf{Notation}: Suppose that $A_S\in\mathbb{R}^{(n-s)\times n}$ is a row submatrix of $A$ containing $(n-s)$ randomly and uniformly chosen rows. $A_i$ (or $A_{S,i}$) denotes $i$th column of matrix $A$ (or $A_S$) and $a_i$ (or $a_{S,i}$) denotes $i$th row of matrix $A$ (or $A_S$). The supp$(x)$ is defined as the support set of vector $x$. $\|x\|_0$ represents the number of nonzero elements in vector $x$.

\section{$0$-Approximate Gradient Code}

In this section, we consider a simplified scenario that the error of gradient evaluation is zero. It can be regarded as a probabilistic relaxation of the worst-case scenario in~\cite{tandon2017gradient}. We first characterize the fundamental limits of the any gradient codes in the set.
\begin{equation*}
\mathcal{G}_{0}=\{A\in\mathbb{R}^{n\times n}|\mathbb{P}[\text{err}(A_S)>0]=o(1)\}.
\end{equation*}
Then we design a gradient code to achieve the lower bound.

\subsection{Minimum Computation Load\label{sec:theory_path}}

The minimum computation load can be determined by exhaustively searching over all possible coding matrices $A\in\mathcal{G}_0$. However, there exist $\Omega(2^{n^2})$ possible candidates in $\mathcal{G}_0$ and such a procedure is practically intractable. To overcome this challenge, we construct a new theoretical path: (i) we first analyze the structure of the optimal gradient codes, and establish a lower bound of the minimum failure probability  $\mathbb{P}(\min\nolimits_{u\in\mathbb{R}^{n-s}}\|A_S^Tu-1_n\|^2>0)$ given computation load $d$; (ii) we derive an exact estimation of such lower bound, which is a monotonically non-increasing function of $d$; and (iii) we show that this lower bound is non-vanishing when the computation $d$ is less than a specific quantity, which provides the desired lower bound.

\begin{figure}[t]
	\vskip -0.1in
	\begin{center}
		\centerline{\includegraphics[width=3.6in]{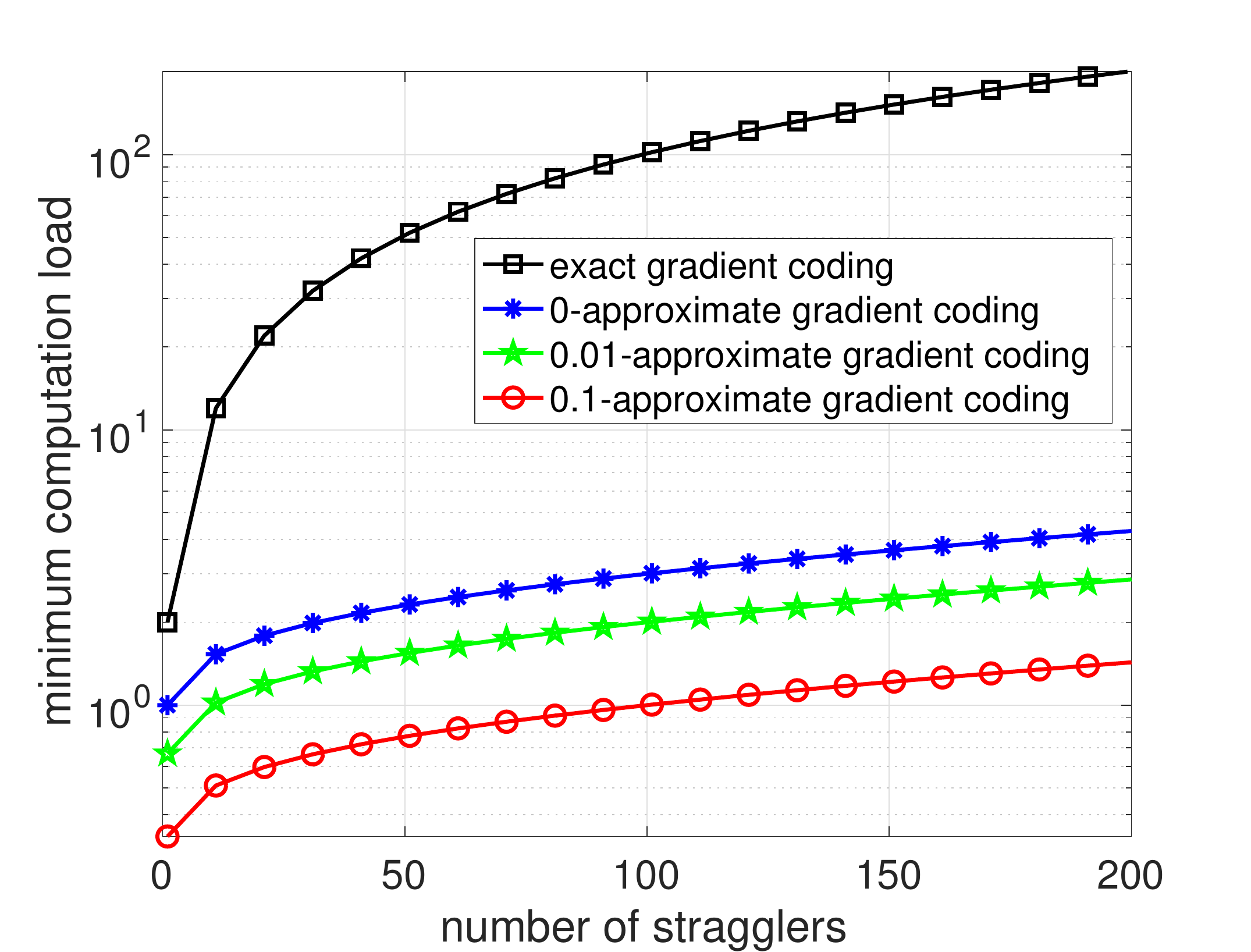}}
		\vskip -0.0in
		\caption{Information-theoretical lower bound of existing worst-case gradient coding~\cite{tandon2017gradient} and proposed $\epsilon$-approximate gradient coding when $n=1000$.}
		\vskip -0.4in
		\label{fig:theory}
	\end{center}
\end{figure}

The following lemma shows that the minimum probability of decoding failure is lower bounded by the minimum probability that there exists an all-zero column of matrix $A_S$ in a specific set of matrices.
\begin{lemma}\label{lm:opt_structure} Suppose that the computation load $\kappa(A)=d$ and define the set of matrices $\mathcal{A}_n^d=\{A\in\mathbb{R}^{n\times n}|\kappa(A)=d\}$, we have
	\begin{equation}\label{eq:exists_zero_proba}
	\min\limits_{A\in\mathcal{A}_n^d}\mathbb{P}(\emph{err}(A_S)>0)\geq \min\limits_{A\in\mathcal{U}_n^d}\mathbb{P}\left(\bigcup\limits_{i=1}^n\|A_{S,i}\|_0=0\right),
	\end{equation}
	where set of matrices $\mathcal{U}_n^d\triangleq\{A\in\mathbb{R}^{n\times n}| \|a_i\|_0=\|A_i\|_0=d,\forall i\in[n]\}$.
\end{lemma}
Based on the inclusion-exclusion principle, we observe that the above lower bound is dependent on the set system $\{\text{supp}(A_{i})\}_{i=1}^n$ formed by matrix $A$. Therefore, one can directly transform the above minimization problem into an integer program. However, due to the non-convexity of the objective function, it is difficult to obtain a closed form expression. To reduce the complexity of our analysis, we have the following lemma to characterize a common structure among all matrices in set $\mathcal{U}_n^d$.
\begin{lemma}\label{lm:independent_set} For any matrix $A\in\mathcal{U}_n^d$, there exists set $\mathcal{I}_d\subseteq[n]$ such that $|\mathcal{I}_d|\geq \lfloor n/d^2\rfloor$ and
	\begin{equation}
	\emph{supp}(A_i)\cap \emph{supp}(A_j)=\emptyset, \forall i\neq j, i,j\in\mathcal{I}_d.
	\end{equation}
\end{lemma}
Based on the results of Lemma~\ref{lm:opt_structure} and Lemma~\ref{lm:independent_set}, we can get an estimation of the lower bound (\ref{eq:exists_zero_proba}), and obtain the following theorem.
\begin{theorem}\label{thm:exact_lower_bound}
	Suppose that out of $n$ workers, $s = \delta n$ are stragglers. If $s=\Omega(1)$, the minimum computation load satisfies
	\begin{equation}
	d^*(s,0)\triangleq\min\limits_{A\in \mathcal{G}_0} \kappa(A) \geq \frac{\log(n\log^2(1/\delta)/\log^{2}(n))}{\log(1/\delta)};
	\end{equation}
	otherwise, $d^*(s,0)=1.$
\end{theorem}
Based on Theorem~\ref{thm:approx_lower_bound}, we can observe the power of probabilistic relaxation in reducing the computation load. For example, if the number of stragglers $s$ is proportional to the number of workers $n$, the minimum computation load $d^*(s,0)=O(\log(n))$, while the worst-case lower bound is $\Theta(n)$; if $s=\theta(n^{\lambda})$, where $0<\lambda<1$ is a constant, the minimum computation load $d^*(s,0)=1/(1-\lambda)$ is a constant, while the worst-case one is $\Theta(n^{\lambda})$. The Figure~\ref{fig:theory} provides a quantitative comparison of the proposed $\epsilon-$approximate gradient coding and existing ones.

\subsection{$d$-Fractional Repetition Code}

In this subsection, we provide a construction of coding matrix $A$ that asymptotically achieves  the minimum computation load. The main idea is based on a generalization of the existing fractional repetition code~\cite{tandon2017gradient}.
\begin{definition}
	\emph{($d$-Fractional Repetition Code)} Divide $n$ workers into $d$ groups of size $n/d$. In each group, divide all data equally and disjointly, and assign $d$ partitions to each worker. All the groups are replicas of each other. The coding matrix $A^{\text{FRC}}$ is defined as
	\begin{equation*}
	A^{\text{FRC}}=\begin{bmatrix}
	A_b\\
	A_b\\
	\vdots\\
	A_b\\
	\end{bmatrix}, A_{b}=\begin{bmatrix}
	1_{1\times d} & 0_{1\times d}  & \cdots & 0_{1\times d}\\
	0_{1\times d} & 1_{1\times d}  & \cdots & 0_{1\times d}\\
	\vdots & \vdots & \ddots & \vdots\\
	0_{1\times d} & 0_{1\times d}  & \cdots & 1_{1\times d}
	\end{bmatrix}_{\frac{n}{d}\times n}.
	\end{equation*}
\end{definition}
Note that we do not need the assumption that $n$ is a multiple $d$. In this case, we can construct the FRC as following: let the size of each group equal to $\lfloor n/d\rfloor$. Then randomly choose mod($n,d$) groups and increase the size of each by one. Besides, the decoding algorithm for the FRC is straightforward: instead of solving the problem (\ref{eq:decode_ls}), the master node sums the partial gradients of any $n/d$ workers that contain disjoint data partitions. The following technical lemma proposed in~\cite{linial1990approximate} is useful in our theoretical analysis.
\begin{lemma}\label{lm:approx_inclusion}
	\emph{(Approximate inclusion-exclusion principle)} Let $n$ be integers and $k\geq \Omega(\sqrt{n})$, and let $E_1, E_2, \ldots,E_n$ be collections of sets, then we have
	\begin{equation*}
	\mathbb{P}\left(\bigcup\limits_{i\in [n]}A_i\right)=\left(1+e^{-\frac{2k}{\sqrt{n}}}\right)\sum\limits_{I\subseteq [n]}^{|I|\leq k}(-1)^{|I|}\cdot\mathbb{P}\left(\bigcap\limits_{i\in I}A_i\right).
	\end{equation*}
\end{lemma}
The above lemma shows that one can approximately estimate the probability of event $E_1\cap\cdots\cap E_n$ given the probability of events $\bigcap\nolimits_{i\in I}E_i, |I|<\Omega(k^{0.5})$.
\begin{theorem}\label{thm:frc_code}
	Suppose that there exist $s=\delta n$ stragglers in $n$ workers. If $d$ satisfies
	\begin{equation}
	d=\max\left\{1,\frac{\log(n\log(1/\delta))}{\log(1/\delta)}\right\},
	\end{equation}
	then we have $\mathbb{P}(\emph{err}(A^{\text{FRC}}_S)>0)=o(1)$.
\end{theorem}
Combining the results of Theorem~\ref{thm:exact_lower_bound} and Theorem~\ref{thm:frc_code}, we can obtain the main argument of Theorem~\ref{thm:main_1}. In practical implementation of FRC, once the decoding process fails in $k$th iteration, a straightforward method is to restart $k$th iteration. Due to the fact that the decoding failure is less happen during the iteration, such overhead will be amortized. As can be seen in the experimental section, during 100 iterations, only one or two iterations are decoding failure.

In this section, we consider a more general scenario that the error of gradient evaluation $\epsilon$ is larger than zero. We first provide a fundamental three-fold trade-off among the computation load, error of gradient and the number of stragglers of any codes in the set.
\begin{equation}
\mathcal{G}_{\epsilon}=\{A\in\mathbb{R}^{n\times n}|\mathbb{P}[\text{err}(A_S)>\epsilon n]=o(1)\}.
\end{equation}
Then we construct a random code that achieves this lower bound.

\subsection{Fundamental Three-fold Tradeoff}

Based on the proposed theoretical path in Section~\ref{sec:theory_path}, we can lower bound the probability that the decoding error is larger than $\epsilon n$ by the one that there exist larger than $\epsilon n$ all-zero columns of matrix $A_S$. However, such a lower bound does not admit a close-form expression since the probability event is complicated and contains exponential many partitions. To overcome this challenge, we decompose the above probability event into $n$ dependent events, and analyze its the second-order moment. Then, we use Bienayme-Chebyshev inequality to further lower bound the above probability. The following theorem provides the lower bound of computation load among the feasible gradient codes $\mathcal{G}_{\epsilon}$.
\begin{theorem}\label{thm:approx_lower_bound}
	Suppose that out of $n$ workers, $s = \delta n$ are stragglers, and $\epsilon<O(1/\log^2(n))$, then the minimum computation load satisfies
	\begin{equation*}
	\kappa^*(A)\triangleq\min\limits_{A\in \mathcal{G}_{\epsilon}} \kappa(A) \geq \frac{\log(n\log^2(1/\delta)/(2\epsilon n+4)\log^{2}(n))}{\log(1/\delta)}.
	\end{equation*}
\end{theorem}
Note that the above result also holds for $\epsilon = 0$, which is slightly lower than the bound in Theorem~\ref{thm:exact_lower_bound}. Based on the result of Theorem~\ref{thm:approx_lower_bound}, we can see that the gradient noise $\epsilon$ provides a logarithmic reduction of the computation load. For example, when $n=1000$ and $s=100$, the $0$-approximate gradient coding requires each worker storing $3\times$ data partitions, while even a  $0.01$-approximate gradient coding only requires $2\times$ data partitions. Detailed comparison can be seen in Figure~\ref{fig:theory}.

\subsection{Random Code Design}

Now we present the construction of our random code, we name \emph{batch rapter code} (BRC), which achieves the above lower bound with high probability. The construction of the BRC consists of two layers. In the first layer, the original data set $\{D_i\}_{i=1}^n$ are partitioned into $n/b$ batches $\{B_i\}_{i=1}^{n/b}$ with the size of each batch equal to $b$. The data in each batch is selected by $B_i=\{D_j\}_{j=1+(i-1)b}^{ib}$, and therefore the intermediate coded partial gradients can be represented by $g^b_i=\sum\nolimits_{j=1+(i-1)b}^{ib}g_j, \forall i\in[n/b]$. In the second step, we construct a type of raptor code taking the coded partial gradients $g^b$ as input block.
\begin{definition}\label{def:brc_code}
	\emph{(($b,P$)-batch rapter code)} Given the degree distribution $P\in\mathbb{R}^{n/b}$ and batches $\{B_i\}_{i=1}^{n/b}$, we define the $(b,P)$-batch rapter code as: each worker $k\in[n]$, stores the data $\{B_i\}_{i\in I}$ and computes
	\begin{equation}
	\tilde{g}_k=\sum\limits_{i\in I} g^b_i=\sum\limits_{i\in I}\sum\limits_{j=1+(i-1)b}^{ib} g_j,
	\end{equation}
	where $I$ is a randomly and uniformly subset of $[n/b]$ with $|I|=d$, and $d$ is generated according to distribution $P$. The coding matrix $A^{\text{BRC}}$ is therefore given by
	
	\begin{equation*}
	A^{\text{BRC}}\triangleq\begin{bmatrix}
	\bovermat{random \text{d} nonzero blocks}{1_{1\times b} & 0_{1\times b}  & 1_{1\times b}} & \cdots & 0_{1\times b}\\
	0_{1\times b} & 0_{1\times b}  & 1_{1\times b} & \cdots & 0_{1\times b}\\
	\vdots & \vdots & \vdots& \ddots & \vdots\\
	0_{1\times b} & 1_{1\times b}  & 0_{1\times b} &\cdots & 1_{1\times b}
	\end{bmatrix}.
	\end{equation*}
\end{definition}
Note that when $n$ is not a multiple $d$, we can tackle it using a method similar to the one used in FRC. The decoding algorithm for the $(b,P)$-batch raptor code goes through a peeling decoding process: it first finds a ripple worker (with only one batch) to recover one batch $g^b_i$ and add it to the sum of gradient $g$. Then for each collected results, it subtracts this batch if the computed gradients contains this batch. The whole procedure is listed in Algorithm~\ref{alg:bpcode}.

\textbf{Example 1.} (Batch raptor code, $n=6,s=2$) Consider a distributed gradient descent problem with $6$ workers. In the batch raptor code, the data is first partitioned into $4$ batches with $B_1=\{D_1\}$, $B_2=\{D_2\}$, $B_3=\{D_3,D_4\}$, $B_4=\{D_5,D_6\}$. After random construction, $6$ workers are assigned the tasks: $\tilde{g_1}=g_1+g_2$, $\tilde{g_2}=g_1$, $\tilde{g_3}=g_2+(g_5+g_6)$, $ \tilde{g_4}=(g_3+g_4)+(g_5+g_6)$, $\tilde{g_5}=g_5+g_6$, $\tilde{g_6}=g_2+(g_5+g_6)$. Suppose that both the $5$th and $6$th workers are stragglers and the master node collects partial results from workers $\{1,2,3,4\}$. Then we can use the peeling decoding algorithm: first, find a ripple node $\tilde{g}_2$ . Then we can use $\tilde{g}_2$ to recover $g_2$ by $\tilde{g}_1-\tilde{g}_2$. Further, we can use $g_2$ to get a new ripple $g_5+g_6$ by $\tilde{g}_3-g_2$, and use ripple $g_5+g_6$ to recover $g_3+g_4$ by $\tilde{g}_4-(g_5+g_6)$. In another case, change the coding scheme of $3$th, $4$th and $6$th worker to $\tilde{g}_3=g_2$, $\tilde{g}_4=g_3+g_4$ and $\tilde{g}_6=g_1+g_2$. Suppose that both the $4$th and $6$th workers are stragglers.  We can use a similar decoding algorithm to recover $g_1+g_2+g_5+g_6$ without $g_3, g_4$. However, the computation load is decreased from $4$ to $2$.
\section{$\epsilon$-Approximate Gradient Code}
\begin{figure*}[t]
	%\vskip 0.1in
	\begin{center}
		\centerline{\includegraphics[width=6.8in]{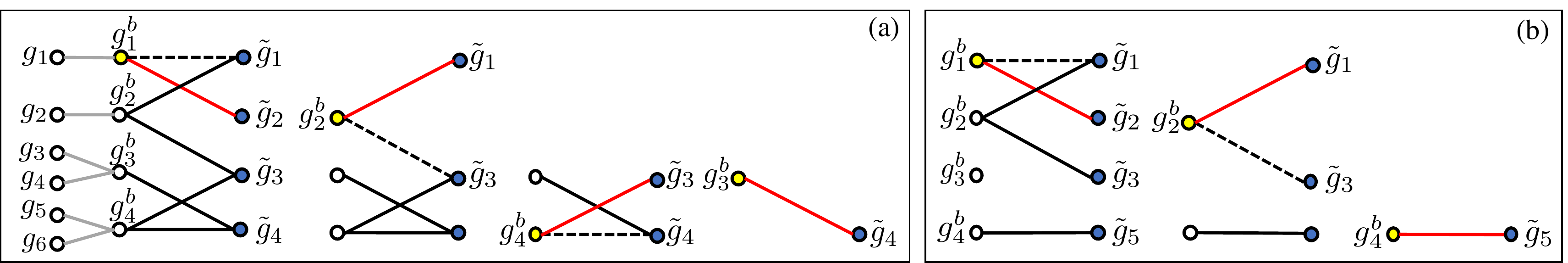}}
		\vskip -0.1in
		\caption{Example of the batch raptor code and peeling decoding algorithm.}
		\label{fig:example}
	\end{center}
	\vskip -0.4in
\end{figure*}

Actually, the above peeling decoding algorithm can be viewed as an edge-removal process in a bipartite graph. We construct a bipartite graph with one partition being the original batch gradients $\{g^b_i\}$ and the other partition being the coded gradients $\{\tilde{g}_i\}$. Two nodes are connected if such a computation task contains that block. As shown in the Figure~\ref{fig:example}, in each iteration, we find a ripple (degree one node) in the right and remove the adjacent edges of that left node, which might produce some new ripples in the right. Then we iterate this process until we decode all gradients. 

Based on the above graphical illustration, the key point of 
being able to successfully decode for $(b,P)$-batch raptor code is the existence of the ripple during the edge removal process, which is mainly dependent on the degree distribution $P$ and batch size $b$. The following theorem shows that, under a specific choice of $P$ and $b$, we can guarantee the success of decoding process with high probability.
\begin{theorem}\label{thm:raptor}
	Define the degree distribution $P_w$
	\begin{equation}\label{eq:soliton}
	p_k=\left\{
	\begin{aligned}
	&\frac{u}{u+1}, k=1;\frac{1}{D(u+1)}, k=D+1\\
	&\frac{1}{k(k-1)(u+1)}, 2\leq k\leq D
	\end{aligned}
	\right.,
	\end{equation}
	where $D=\lfloor 1/\epsilon \rfloor$, $u=2\epsilon(1-2\epsilon)/(1-4\epsilon)^2$ and $b=\lceil 1/\log(1/\delta)\rceil+1$. Then the $(\lceil 1/\log(1/\delta)\rceil,P_w)$-batch rapter code with decoding Algorithm~\ref{alg:bpcode} satisfies
	\begin{equation*}
	\mathbb{P}(\emph{err}(A^{\text{BRC}}_S)>c)<e^{-c_0n},
	\end{equation*}
	and achieves an average computation load of
	\begin{equation}
	O\left(\frac{\log(1/\epsilon)}{\log(1/\delta)}\right).
	\end{equation}
\end{theorem}
The above result is based on applying a martingale argument to the peeling decoding process~\cite{luby2001efficient}. In practical implementation, the degree distribution can be further optimized given the $n$, $s$ and error $\epsilon$~\cite{shokrollahi2006raptor}.

\begin{algorithm}[htb]
	\caption{Batch raptor code (master node's protocol)}
	\label{alg:bpcode}
	\begin{algorithmic}
		\REPEAT
		\STATE The master node assign the data sets according to Definition~\ref{def:brc_code}.
		\UNTIL{the master node collects results from first finished $n-s$ workers}.
		\REPEAT
		\STATE Find a row $M_{i}$ in received coding matrix $M$ with $\|M_i\|_0=1$.
		\STATE Suppose that the column index of the nonzero element in matrix $M_i$ is $k_0$ and let $g=g+\tilde{g}_{k_0}$.
		\FOR{each computation results $\tilde{g}_{k}$}
		\IF{$M_{kk_0}$ is nonzero}
		\STATE $\tilde{g}_{k}=\tilde{g}_{k}-M_{kk_0}\tilde{g}_{k_0}$ and set $M_{kk_0}=0$.
		\ENDIF
		\ENDFOR
		\UNTIL{$n(1-\epsilon)$ partial gradients is recovered.}
	\end{algorithmic}
\end{algorithm}

\section{Simulation Results}

In this section, we present the experimental results at Ohio Supercomputer center~\cite{OhioSupercomputerCenter1987}. We compare our proposed schemes including $d$-fractional repetition code (FRC) and batch raptor code (BRC) against existing gradient coding schemes: (i) forget-$s$ scheme (stochastic gradient descent): the master node only waits the results of non-straggling workers; (ii) cyclic MDS code~\cite{tandon2017gradient}: gradient coding scheme that can guarantee the decodability for any $s$ stragglers; (iii) bernoulli gradient code (BGC)~\cite{charles2017approximate}: approximate gradient coding scheme that only requires $O(\log(n))$ data copies in each worker. To simulate straggler effects in a large-scale system, we randomly pick $s$ workers that are running a background thread.

\subsection{Experiment Setup}
We implement all methods in python using MPI4py. Each worker stores the data according to the coding matrix $A$. During the iteration of the distributed gradient descent, the master node broadcasts the current classifier $\beta^{(t)}$ using \texttt{Isend()}; then each worker computes the coded partial gradient $\tilde{g}^{(t)}_i$ and returns the results using \texttt{Isend()}. Then the master node actively listens to the response from each worker via \texttt{Irecv()}, and uses \texttt{Waitany()} to keep polling for the earliest finished tasks. Upon receiving enough results, the master stops listening and starts decoding the full gradient $g^{(t)}$ and updates the classifier to $\beta^{(t+1)}$.

In our experiment, we ran various schemes to train logistic regression models, a well-understood convex optimization problem that is widely used in practice. We choose the training data from LIBSVM dataset repository. We use $N=19264097$ samples and a model dimension of $p=1163024$. We evenly divide the data into $n$ partitions $\{D_k\}_{k=1}^{n}$. The key step of gradient descent algorithm is
\begin{equation*}
\beta^{(t+1)}=\beta^{(t)}+\alpha\sum\limits_{k=1}^{n}\overbrace{\sum\limits_{i\in D_k}\eta(y_i-h_{\beta^{(t)}}(x_i))x_i}^{g_{i}^{(t)}},
\end{equation*}
where $h_{\beta^{(t)}}(\cdot)$ is the logistic function, $\alpha$ is the predetermined step size.

\subsection{Generalization Error}

\begin{figure*}[t]
	\vskip -0.0in
	\begin{center}
		\centerline{\includegraphics[width=6.6in]{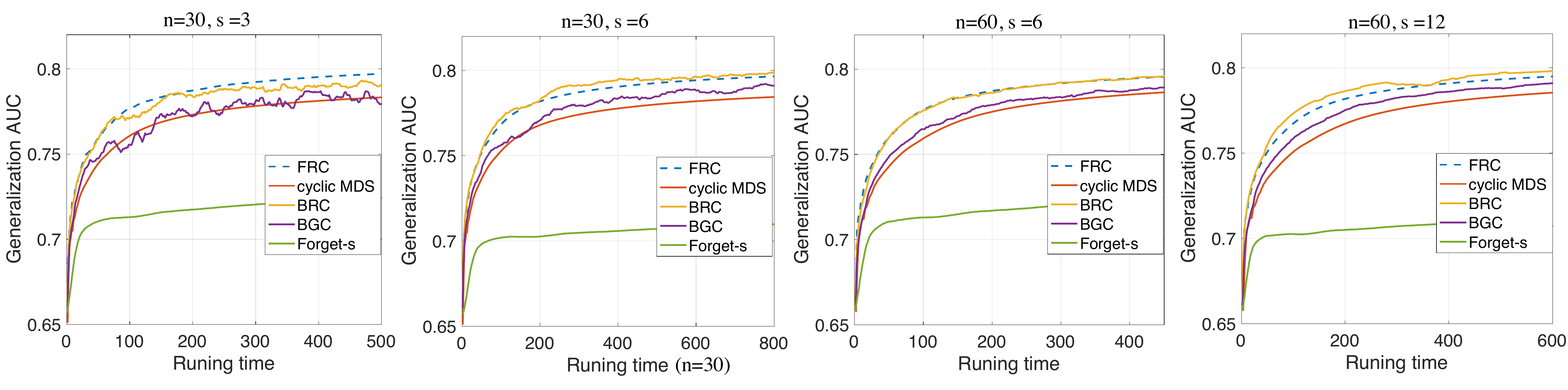}}
		\vskip -0.15in
		\caption{The generalization AUC versus running time of applying distributed gradient descent in a logistic regression model. The two proposed schemes FRC and BRC are compared against three existing schemes. The learning rate $\alpha$ is fixed for all the experiment.}
		\vskip -0.2in
		\label{fig:simres_Auc}
	\end{center}
\end{figure*}

We first compare the generalization AUC of the above five schemes when number of workers $n=30$ or $60$ and $10\%$ or $20\%$ workers are stragglers. In Figure~\ref{fig:simres_Auc}, we plot the generalization AUC versus the running time of all the schemes under different $n$ and $s$. We can observe that our proposed schemes (FRC and BRC) achieve significantly better generalization error compared to existing ones. The forget-$s$ scheme (stochastic gradient descent) converges slowly, since it does not utilize the full gradient and only admits a small step size $\alpha$ compared to other schemes. In particular, when the number of workers increases, our proposed schemes provide even larger speed up over the state of the art. 

\subsection{Impact of Straggler Tolerance}

\begin{figure*}[t]
	\vskip -0.1in
	\begin{center}
		\centerline{\includegraphics[width=6.6in]{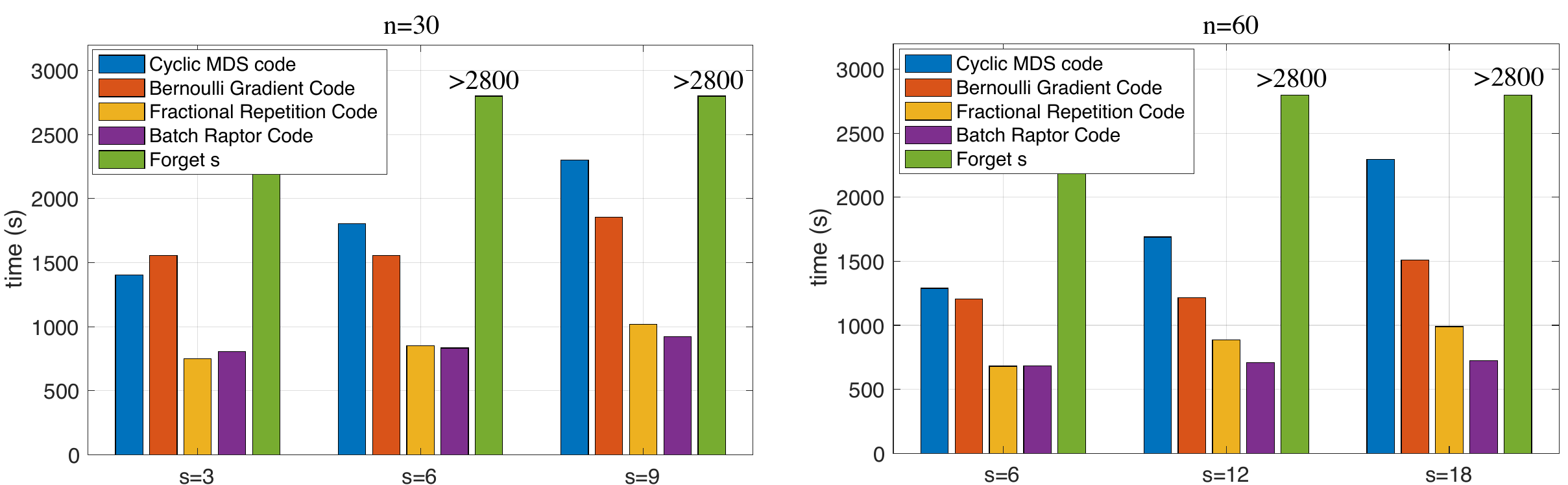}}
		\vskip -0.0in
		\caption{The final job completion time of achieving a generalization AUC $=0.8$ in a logistic regression model. The two proposed schemes FRC and BRC are compared against three existing schemes. The learning rate $\alpha$ is fixed for all the experiments.}
		\vskip -0.3in
		\label{fig:simres_tot}
	\end{center}
\end{figure*}

We further investigate the impact of straggler tolerance $s$. We fix the number of workers $n=30$ or $n=60$ and increase the fraction of stragglers from $10\%$ to $30\%$. In Figure~\ref{fig:simres_tot}, we plot the job completion time to achieve a fixed generalization AUC $=0.8$. The first observation is that our propose schemes reduce the completion time by $50\%$ compared to existing ones. The cyclic MDS code and forget-$s$ (stochastic gradient descent) schemes are sensitive to the number of stragglers. The main reasons are: (i) the computation load of cyclic MDS code is linear in $s$; (ii) the available step size of forget-$s$ scheme is reduced when the number of received partial gradients decreases. 

The job completion time of the proposed FRC and BRC are not sensitive to the straggler tolerance, especially when the number of workers $n$ is large. For example, the completion time of batch raptor code only increases $10\%$ when fraction of straggler increases from $0.1$ to $0.3$. Besides, we observe that, when the straggler tolerance is small, i.e., $s/n<0.1$, the FRC is slightly better than the BRC, since the computation loads are similar in this case and the FRC utilizes the information of the full gradient.

%Therefore, we suggest that using the fractional repetition code when the straggler tolerance is small, i.e., $s/n<0.1$, while using the approximate gradient coding scheme batch raptor code when the straggler tolerance is large, i.e., $s/n>0.1$ from $0.1$ to $0.3$. 

\section{Conclusion}

In this paper, we formalized the problem of approximate gradient coding and systematically characterized the fundamental three-fold tradeoff among the computation load, noise of gradient $\epsilon$ and straggler tolerance $s$. We further constructed two schemes, FRC and BRC, to achieve this lower bound. Theoretically, the proposed fundamental tradeoff uncovers how the ``probabilistic relaxation'' and the ``noise of gradient'' quantitatively influence the computation load: (i) the probabilistic relaxation provides a linear reduction in computation load, i.e, a reduction from $\Theta(n)$ to $O(\log(n))$ redundant computations of each worker; (ii) the gradient noise $\epsilon$ introduces another logarithmic reduction in computation load, i.e, a reduction from $O(\log(n))$ to $O(\log(n)-\log(\epsilon n))$ redundant computations. In practice, we have experimented with various gradient coding schemes on super computing center. Our proposed schemes provide $50\%$ speed up over the state of the art.

\bibliography{example_paper}

\begin{thebibliography}{10}
\providecommand{\url}[1]{#1}
\csname url@samestyle\endcsname
\providecommand{\newblock}{\relax}
\providecommand{\bibinfo}[2]{#2}
\providecommand{\BIBentrySTDinterwordspacing}{\spaceskip=0pt\relax}
\providecommand{\BIBentryALTinterwordstretchfactor}{4}
\providecommand{\BIBentryALTinterwordspacing}{\spaceskip=\fontdimen2\font plus
\BIBentryALTinterwordstretchfactor\fontdimen3\font minus
  \fontdimen4\font\relax}
\providecommand{\BIBforeignlanguage}[2]{{%
\expandafter\ifx\csname l@#1\endcsname\relax
\typeout{** WARNING: IEEEtran.bst: No hyphenation pattern has been}%
\typeout{** loaded for the language `#1'. Using the pattern for}%
\typeout{** the default language instead.}%
\else
\language=\csname l@#1\endcsname
\fi
#2}}
\providecommand{\BIBdecl}{\relax}
\BIBdecl

\bibitem{tandon2017gradient}
R.~Tandon, Q.~Lei, A.~G. Dimakis, and N.~Karampatziakis, ``Gradient coding:
  Avoiding stragglers in distributed learning,'' in \emph{International
  Conference on Machine Learning}, 2017, pp. 3368--3376.

\bibitem{le2011optimization}
Q.~V. Le, J.~Ngiam, A.~Coates, A.~Lahiri, B.~Prochnow, and A.~Y. Ng, ``On
  optimization methods for deep learning,'' in \emph{Proceedings of the 28th
  International Conference on International Conference on Machine
  Learning}.\hskip 1em plus 0.5em minus 0.4em\relax Omnipress, 2011, pp.
  265--272.

\bibitem{dean2008mapreduce}
J.~Dean and S.~Ghemawat, ``Mapreduce: simplified data processing on large
  clusters,'' \emph{Communications of the ACM}, vol.~51, no.~1, pp. 107--113,
  2008.

\bibitem{zaharia2010spark}
M.~Zaharia, M.~Chowdhury, M.~J. Franklin, S.~Shenker, and I.~Stoica, ``Spark:
  Cluster computing with working sets.'' \emph{HotCloud}, vol.~10, no. 10-10,
  p.~95, 2010.

\bibitem{dean2012large}
J.~Dean, G.~Corrado, R.~Monga, K.~Chen, M.~Devin, M.~Mao, A.~Senior, P.~Tucker,
  K.~Yang, Q.~V. Le \emph{et~al.}, ``Large scale distributed deep networks,''
  in \emph{Advances in neural information processing systems}, 2012, pp.
  1223--1231.

\bibitem{yadwadkar2016multi}
N.~J. Yadwadkar, B.~Hariharan, J.~E. Gonzalez, and R.~Katz, ``Multi-task
  learning for straggler avoiding predictive job scheduling,'' \emph{The
  Journal of Machine Learning Research}, vol.~17, no.~1, pp. 3692--3728, 2016.

\bibitem{dutta2016short}
S.~Dutta, V.~Cadambe, and P.~Grover, ``Short-dot: Computing large linear
  transforms distributedly using coded short dot products,'' in \emph{Advances
  In Neural Information Processing Systems}, 2016, pp. 2100--2108.

\bibitem{raviv2017gradient}
N.~Raviv, I.~Tamo, R.~Tandon, and A.~G. Dimakis, ``Gradient coding from cyclic
  mds codes and expander graphs,'' 2018.

\bibitem{bottou2010large}
L.~Bottou, ``Large-scale machine learning with stochastic gradient descent,''
  in \emph{Proceedings of COMPSTAT'2010}.\hskip 1em plus 0.5em minus
  0.4em\relax Springer, 2010, pp. 177--186.

\bibitem{neelakantan2015adding}
A.~Neelakantan, L.~Vilnis, Q.~V. Le, I.~Sutskever, L.~Kaiser, K.~Kurach, and
  J.~Martens, ``Adding gradient noise improves learning for very deep
  networks,'' \emph{arXiv preprint arXiv:1511.06807}, 2015.

\bibitem{charles2017approximate}
Z.~Charles, D.~Papailiopoulos, and J.~Ellenberg, ``Approximate gradient coding
  via sparse random graphs,'' \emph{arXiv preprint arXiv:1711.06771}, 2017.

\bibitem{OhioSupercomputerCenter1987}
O.~S. Center, ``Ohio supercomputer center,''
  \url{http://osc.edu/ark:/19495/f5s1ph73}, 1987.

\bibitem{lee2017speeding}
K.~Lee, M.~Lam, R.~Pedarsani, D.~Papailiopoulos, and K.~Ramchandran, ``Speeding
  up distributed machine learning using codes,'' \emph{IEEE Transactions on
  Information Theory}, 2017.

\bibitem{lee2017high}
K.~Lee, C.~Suh, and K.~Ramchandran, ``High-dimensional coded matrix
  multiplication,'' in \emph{Information Theory (ISIT), 2017 IEEE International
  Symposium on}.\hskip 1em plus 0.5em minus 0.4em\relax IEEE, 2017, pp.
  2418--2422.

\bibitem{yu2017polynomial}
Q.~Yu, M.~Maddah-Ali, and S.~Avestimehr, ``Polynomial codes: an optimal design
  for high-dimensional coded matrix multiplication,'' in \emph{Advances in
  Neural Information Processing Systems}, 2017, pp. 4406--4416.

\bibitem{wang2018}
S.~Wang, J.~Liu, and N.~Shroff, ``Coded sparse matrix multiplication,'' in
  \emph{International Conference on Machine Learning}, 2018.

\bibitem{wang2019}
S.~Wang, J.~Liu, N.~Shroff, and P.~Yang, ``Computation efficient coded linear
  transform,'' in \emph{International Conference on Artificial Intelligence and
  Statistics}, 2019.

\bibitem{karakus2017straggler}
C.~Karakus, Y.~Sun, S.~Diggavi, and W.~Yin, ``Straggler mitigation in
  distributed optimization through data encoding,'' in \emph{Advances in Neural
  Information Processing Systems}, 2017, pp. 5434--5442.

\bibitem{maity2018robust}
R.~K. Maity, A.~S. Rawat, and A.~Mazumdar, ``Robust gradient descent via moment
  encoding with ldpc codes,'' \emph{SysML}, 2018.

\bibitem{ye18}
M.~Ye and E.~Abbe, ``Communication-computation efficient gradient coding,'' in
  \emph{International Conference on Machine Learning}, 2018.

\bibitem{linial1990approximate}
N.~Linial and N.~Nisan, ``Approximate inclusion-exclusion,''
  \emph{Combinatorica}, vol.~10, no.~4, pp. 349--365, 1990.

\bibitem{luby2001efficient}
M.~G. Luby, M.~Mitzenmacher, M.~A. Shokrollahi, and D.~A. Spielman, ``Efficient
  erasure correcting codes,'' \emph{IEEE Transactions on Information Theory},
  vol.~47, no.~2, pp. 569--584, 2001.

\bibitem{shokrollahi2006raptor}
A.~Shokrollahi, ``Raptor codes,'' \emph{IEEE/ACM Transactions on Networking
  (TON)}, vol.~14, no.~SI, pp. 2551--2567, 2006.

\end{thebibliography}
\bibliographystyle{IEEEtran}

\appendix
\subsection{Proof of Lemma~\ref{lm:opt_structure}}
\begin{proof}
	Since the event that there exists $i\in[n]$ such that $\|A_{S,i}\|_0=0$ implies the event that err$(A_S)=\|A_S^Tu-1_n\|^2\geq 1$, we can obtain
	\begin{equation*}
	\mathbb{P}(\text{err}(A_S)>0)\geq
	\mathbb{P}\left(\bigcup\limits_{i=1}^n\|A_{S,i}\|_0=0\right)
	\end{equation*}
	Suppose that $A^*=\argmin_{A\in \mathcal{A}_n^d}	\mathbb{P}(\text{err}(A_S)>0)$, $A_S^*$ is the row submatrix of $A^*$ containing $(n-s)$ randomly and uniformly chosen rows, we have
	\begin{align*}
	\min_{A\in\mathcal{A}_n^d}\mathbb{P}(\text{err}(A_S)>0)&\geq
	\mathbb{P}\left(\bigcup\limits_{i=1}^n\|A^*_{S,i}\|_0=0\right)\\
	&\geq \min\limits_{A\in\mathcal{A}_n^d}\mathbb{P}\left(\bigcup\limits_{i=1}^n\|A_{S,i}\|_0=0\right)
	\end{align*}
	We next show that $\min\nolimits_{A\in\mathcal{A}_n^d}\mathbb{P}\left(\bigcup\nolimits_{i=1}^n\|A_{S,i}\|_0=0\right)=\min\nolimits_{A\in\mathcal{U}_n^d}\mathbb{P}\left(\bigcup\nolimits_{i=1}^n\|A_{S,i}\|_0=0\right)$. We will prove that the above probability is monotonically decreasing with the support size of each row and column of matrix $A$. Assume that there exists $k\in[n]$ such that $\|a_k\|_0<\kappa(A)$.  We change one zero position of row $a_k$, i.e., $a_{kj}$ to an nonzero constant. Define the new matrix as $A'$. For simplicity, define the event $E_i$ as $\|A_{S,i}\|_0=0$ and $E_i'$ as $\|A_{S,i}'\|_0=0$. Then we can write
	\begin{align*}
	\mathbb{P}\left(\bigcup\limits_{i=1}^nE_i\right)&=\mathbb{P}\left(\bigcup\limits_{i\neq j}E_i\right)+\mathbb{P}\left[E_j\backslash\left(\bigcup\limits_{i\neq j}E_i\right)\right]\notag\\
	&\overset{(a)}{\geq}\mathbb{P}\left(\bigcup\limits_{i\neq j}E_i'\right)+\mathbb{P}\left[E_j'\backslash\left(\bigcup\limits_{i\neq j}E_i'\right)\right]\\
	&=\mathbb{P}\left(\bigcup\limits_{i=1}^nE_i'\right).
	\end{align*}
	The above, step (a) is based on the fact that $E_i=E_i',i\neq j$ and  $E_j'\subset E_j$. Similarly, we can prove the monotonicity for support size of each column. Therefore, based on the monotonicity and the Definition~\ref{def:computation_load} of computation load, the lemma follows.
\end{proof}

\subsection{Proof of Lemma~\ref{lm:independent_set}}
\begin{proof}
	Given any matrix $A\in\mathcal{U}_n^d$. We construct the set $\mathcal{I}_d$ as follows. First, choose a column $A_{i_1}$ and construct set $I_1$ as follows,
	\begin{equation}
	I_1=\{j\in[n]|\text{supp}(A_{i_1})\cap \text{supp}(A_{j})\neq\emptyset,j\neq i_1\}
	\end{equation}
	Since $A\in\mathcal{U}_n^d$, suppose that $\text{supp}(A_{i_1})=\{k_1,k_2,\ldots,k_d\}$. We can obtain
	\begin{align}
	|I_1|&=\left|\bigcup\limits_{l=1}^d\left\{j\in[n]|A_{jk_i},j\neq i_1\right\}\right|
	\notag\\
	&\overset{(a)}{\leq}\sum\limits_{i=1}^d|\left\{j\in[n]|A_{jk_i},j\neq i_1\right\}|\overset{(b)}{\leq} d^2.
	\end{align}
	The above, step (a) utilizes the union bound and step (b) is based on the definition of set $\mathcal{U}_n^d$. 
	
	Furthermore, we choose a column $A_{i_2}$ such that $i_2\in[n]\backslash I_1$. Based on the definition of index set $I_1$, we have $\text{supp}(A_{i_2})\cap \text{supp}(A_{i_1})=\emptyset$. Similarly, we can construct the index set $I_2=\{j\in[n]|\text{supp}(A_{i_2})\cap \text{supp}(A_{j})\neq\emptyset,j\neq i_1\}$ with $|I_2|\leq d^2$. Continue this process $k$ times, we can construct a set $\mathcal{I}_d=\{i_1,i_2,\ldots,i_k\}$ such that for any $i,j\in\mathcal{I}_d$, $\text{supp}(A_i)\cap \text{supp}(A_j)=\emptyset$, and corresponding $I_1,I_2,\ldots,I_k$. Since each $|I_k|\leq d^2$, we have $|\mathcal{I}_d|\geq \lfloor n/d^2\rfloor$.
\end{proof}

\subsection{Proof of Theorem~\ref{thm:exact_lower_bound}}
\begin{proof}
	Suppose that 
	\begin{equation}
	A^*=\argmin\limits_{A\in \mathcal{U}_n^d}\mathbb{P}\left(\bigcup\limits_{i=1}^n\|A_{S,i}\|_0=0\right).
	\end{equation}
	Based on the results of Lemma~\ref{lm:independent_set}, we can construct a set $I_d^*$ such that $|\mathcal{I}_d^*|\geq \lfloor n/d^2\rfloor$ and 
	\begin{equation}
	\text{supp}(A_i^*)\cap \text{supp}(A_j^*)=\emptyset, \forall i\neq j, i,j\in\mathcal{I}_d.
	\end{equation}
	Combining the results of Lemma~\ref{lm:opt_structure}, we have
	\begin{align}
	\min\limits_{A\in\mathcal{A}_n^d}\mathbb{P}(\text{err}(A_S)>0)&\geq \min\limits_{A\in\mathcal{A}_n^d}\mathbb{P}\left(\bigcup\limits_{i=1}^n\|A_{S,i}\|_0=0\right)\notag\\
	&= \min\limits_{A\in \mathcal{U}_n^d}\mathbb{P}\left(\bigcup\limits_{i=1}^n\|A_{S,i}\|_0=0\right)\notag\\
	&= \mathbb{P}\left(\bigcup\limits_{i=1}^n\|A_{S,i}^*\|_0=0\right)\notag\\
	&\geq \mathbb{P}\left(\bigcup\limits_{i\in I_d^*}\|A_{S,i}^*\|_0=0\right)
	\end{align}
	The above, last step is based on the fact that $I_d^*\subseteq [n]$. Suppose that $|I_d^*|=t$. Based on the inclusion-exclusion principle, we can write
	\begin{align}
	&\mathbb{P}\left(\bigcup\limits_{i\in I_d^*}\|A_{S,i}^*\|_0=0\right)\notag\\
	=&\sum\limits_{I\subseteq I_d^*}(-1)^{|I|+1}\mathbb{P}\left(\bigcap\limits_{i\in I}\|A_{S,i}^*\|_0=0\right)\notag\\
	\overset{(a)}{=}&\sum\limits_{I\subseteq I_d^*,|I|\leq\lfloor \frac{s}{d}\rfloor}(-1)^{|I|+1}\binom{n-\sum\nolimits_{i\in I}|\text{supp}(A_{S,i}^*)|}{s-\sum\nolimits_{i\in I}|\text{supp}(A_{S,i}^*)|}\bigg/\binom{n}{s}\notag\\
	=&\sum\limits_{k=1}^{\min\{t,\lfloor \frac{s}{d}\rfloor\}}\binom{t}{k}(-1)^{k+1}\binom{n-kd}{s-kd}\bigg/\binom{n}{s}\notag\\
	\overset{(b)}{\geq}&\sum\limits_{k\text{ is odd}}^{k\leq \min\{t,\lfloor \frac{s}{d}\rfloor\}}\binom{t}{k} \left(\frac{n-kd}{n}\right)^{n-kd+0.5}\left(\frac{s}{s-kd}\right)^{s-kd+0.5}\left(\frac{s}{n}\right)^{kd}-\notag\\
	&\sum\limits_{k\text{ is even}}^{k\leq \min\{t,\lfloor \frac{s}{d}\rfloor\}}\binom{t}{k}\frac{144s(n-kd)}{(12s-1)(12(n-kd)-1)} \left(\frac{n-kd}{n}\right)^{n-kd+0.5}\left(\frac{s}{s-kd}\right)^{s-kd+0.5}\left(\frac{s}{n}\right)^{kd}.
	\end{align}
	The above, step (a) is based on the property of set $I_d^*$. Step (b) utilizes the following Sterlin's inequalities
	\begin{equation}\label{eq:sterlin}
	\sqrt{2\pi n}\left(\frac{n}{e}\right)^n\leq n!\leq\sqrt{2\pi n}\left(\frac{n}{e}\right)^n\frac{12n}{12n-1}.
	\end{equation}
	\textbf{Case 1: } The number of stragglers $s=\delta n$ and $\delta=\Theta(1)$.
	
	Since the event $\|A_{S,i_0}^*\|_0=0$ belongs to event $\bigcup\nolimits_{i\in I_d^*}\|A_{S,i}^*\|_0=0$ for some $i_0\in I_d^*$, we have
	\begin{align}
	\mathbb{P}\left[\bigcup\limits_{i\in I_d^*}\|A_{S,i}^*\|_0=0\right]\geq&\mathbb{P}\left[ \|A_{S,i_0}^*\|_0=0\right]=\binom{n-d}{s-d}\bigg/\binom{n}{s}\notag\\
	\overset{(a)}{=}&(1+o(1))\left(1-\frac{d}{n}\right)^{n-d+0.5}\left(1+\frac{d}{s-d}\right)^{s-d+0.5}\delta^d\notag\\
	=&(1+o(1))\delta^d.
	\end{align}
	The above, step (a) is based on the Sterlin's approximation. This result implies $d>\Omega(1)$ (otherwise, the failure probability is nonvanishing). Then we have $t=\lfloor n/d^2\rfloor<\lfloor s/d\rfloor$ and $kd=o(n)$ for any $1\leq k\leq t$, and obtain the following approximation,
	\begin{equation*}
	\left(\frac{n-kd}{n}\right)^{n-kd+0.5}\left(\frac{s}{s-kd}\right)^{s-kd+0.5}=1-\alpha(n,k) \text{ and } \lim\limits_{n\rightarrow \infty}\alpha(n,k) = 0, \forall 1\leq k \leq t.
	\end{equation*}
	\begin{equation*}
	\frac{144s(n-kd)}{(12s-1)(12(n-kd)-1)}=1+\beta(n,k) \text{ and } \lim\limits_{n\rightarrow \infty}\beta(n,k) = 0,  \forall 1\leq k \leq t.
	\end{equation*}
	Utilizing the above approximation and choosing $d$ such that $\delta^dt\rightarrow 1/e$, we have
	\begin{align}
	&\mathbb{P}\left(\bigcup\limits_{i\in I_d^*}\|A_{S,i}^*\|_0=0\right)\notag\\
	\geq&\sum\limits_{k\text{ is odd}}^{k\leq t}\binom{t}{k} \left(1-\alpha(n,k)\right)\left(\frac{s}{n}\right)^{kd}-\sum\limits_{k\text{ is even}}^{k\leq t}\binom{t}{k} \left(1-\alpha(n,k)\right)\left(1+\beta(n,k)\right)\left(\frac{s}{n}\right)^{kd}\notag\\
	=&1-(1-\delta^d)^t-\sum\limits_{k\text{ is odd}}^{k\leq t}\binom{t}{k} \alpha(n,k)\delta^{kd}+\sum\limits_{k\text{ is even}}^{k\leq t}\binom{t}{k} (\alpha(n,k)-\beta(n,k)+\alpha(n,k)\beta(n,k))\delta^{kd}\notag\\
	\overset{(a)}{=}&1-(1-\delta^d)^t+o(1)\notag\\
	\overset{(b)}{=}&1-e^{-e^{-1}}+o(1)>0.307.
	\end{align}
	The above, step (a) utilizes the fact that $\binom{t}{k}\leq (et/k)^k$, then the quantity $\binom{t}{k}\delta^{kd}\leq (et\delta^d/k)^k=1/k^k$ and
	\begin{equation}\label{eq:same_pf_1}
	\sum\limits_{k=1}^t\binom{t}{k}(-1)^{k+1} o(1)\delta^{kd}\leq \sum\limits_{k=1}^t \frac{o(1)}{k^k}=o(1).
	\end{equation}
	Step (b) is based on the choice of $d$ such that $\delta^d t\rightarrow 1/e$. It is obvious that the probability $\mathbb{P}(\bigcup\nolimits_{i\in I_d^*}\|A_{S,i}^*\|_0=0)$ is monotonically non-increasing with the computation load $d$. Therefore, the minimum computation load $d^*$ should satisfy $d^*>d_0$, where $\delta^{d_0}n/d_0^2\rightarrow 1/e$. It is easy to see that
	\begin{equation}\label{eq:d_0_exact_lower_bound}
	d_0=\frac{\log(ne\log^2(1/\delta)/\log^2(n))}{\log(1/\delta)}.
	\end{equation}
	
	\textbf{Case 2: } The number of stragglers $s=\delta n$, $\delta=o(1)$ and $\delta=\Omega(1/n)$.
	
	In this case, we can choose $d=d_0$. The conditions $\delta=\Omega(1/n)$ implies that $s=\Omega(1)$. 
	
	Then, for $k\in[\min\{t,s/d_0\}]$ and $k=o(s/d_0)$, we have following similar estimation.
	\begin{equation}\label{eq:pf:estimation_1}
	\left(\frac{n-kd_0}{n}\right)^{n-kd_0+0.5}\left(\frac{s}{s-kd_0}\right)^{s-kd_0+0.5}=1-\alpha(n,k) \text{ and } \lim\limits_{n\rightarrow \infty}\alpha(n,k) = 0.
	\end{equation}
	For $k\in[\min\{t,s/d_0\}]$ and $k=\Theta(s/d_0)=cs/d_0$, we have 
	\begin{align}
	\binom{t}{k}\left(\frac{n-kd_0}{n}\right)^{n-kd_0+0.5}\left(\frac{s}{s-kd_0}\right)^{s-kd_0+0.5}\delta^{kd_0}=&
	\binom{t}{k}\left(\frac{1}{1-c}\right)^{(1-c)s+0.5}\delta^{cs}\notag\\
	\leq&\left(\frac{1}{1-c}\right)^{0.5}\left[\left(\frac{e}{c}\right)^{1/d_0}\left(\frac{1}{1-c}\right)^{1/c-1}\delta\right]^{kd_0}.\label{eq:pf:estimation_2}
	\end{align}
	For all $k\in[\min\{t,s/d_0\}]$, we have
	\begin{equation}\label{eq:pf:estimation_3}
	\frac{144s(n-kd_0)}{(12s-1)(12(n-kd_0)-1)}=1+\beta(n,k) \text{ and } \lim\limits_{n\rightarrow \infty}\beta(n,k) = 0,  \forall 1\leq k \leq t.
	\end{equation}
	Therefore, we can obtain the following estimation.
	\begin{equation}\label{eq:pf:estimation_4}
	\sum\limits_{k=\Theta(s/d_0)}\binom{t}{k}\left(\frac{n-kd_0}{n}\right)^{n-kd_0+0.5}\left(\frac{s}{s-kd_0}\right)^{s-kd_0+0.5}\delta^{kd_0}=o(1).
	\end{equation}
	Utilizing the above approximation, we have
	\begin{align}
	&\mathbb{P}\left(\bigcup\limits_{i\in I_{d_0}^*}\|A_{S,i}^*\|_0=0\right)\notag\\
	\geq&\sum\limits_{k\text{ is odd}}^{k=o(s/d_0)}\binom{t}{k} \left(1-\alpha(n,k)\right)\left(\frac{s}{n}\right)^{kd_0}-\sum\limits_{k\text{ is even}}^{k=o(s/d_0)}\binom{t}{k} \left(1-\alpha(n,k)\right)\left(1+\beta(n,k)\right)\left(\frac{s}{n}\right)^{kd_0}+o(1)\notag\\
	=&1-(1-\delta^{d_0})^t-\sum\limits_{k\text{ is odd}}^{k=o(s/d_0)}\binom{t}{k} \alpha(n,k)\delta^{kd_0}+\sum\limits_{k\text{ is even}}^{k=o(s/d_0)}\binom{t}{k} (\alpha(n,k)-\beta(n,k)+\alpha(n,k)\beta(n,k))\delta^{kd_0}+o(1)\notag\\
	=&1-(1-\delta^{d_0})^t+o(1)\notag\\
	=&1-e^{-e^{-1}}+o(1)>0.307.
	\end{align}
	Therefore, the minimum computation load $d^*$ should satisfy $d^*>d_0$. In the case $s=\Theta(1)$, the lower bound $1$ is trivial (otherwise, some gradients are lost).Therefore, the theorem follows.
\end{proof}

\subsection{Proof of Theorem~\ref{thm:frc_code}}

\begin{proof}
	Based on the structure of coding matrix $A^{\text{FRC}}$ and decoding algorithm, we can define the following event
	\begin{equation}
	E_i \triangleq \bigcap\limits_{j=0}^{d-1}\left(\frac{jn}{d}+i\right) \text{th worker }  \text{is straggler}, 1\leq i\leq n/d,
	\end{equation}
	and we have
	\begin{align}
	&\mathbb{P}(\text{err}(A^{\text{FRC}}_S)>0) = \mathbb{P}\left(\bigcup\limits_{i=1}^{n/d} E_i\right).
	\end{align}
	Utilizing the approximate inclusion-exclusion principle and choose $k=n^{0.6}$ and $d=1+\log(n)/\log(1/\delta)$, we have
	\begin{align}
	&\mathbb{P}(\text{err}(A^{\text{FRC}}_S)>0) = (1+e^{-2n^{0.1}}) \sum\limits_{I\subseteq[n],|I|\leq k} (-1)^{|I|+1}\mathbb{P}\left(\bigcap\limits_{i\in I} E_i\right)\notag\\
	\overset{(a)}{=}&(1+o(1))\sum\limits_{i=1}^{k}\binom{n/d}{i} (-1)^{i+1}\mathbb{P}\left(\bigcap\limits_{j=1}^{i} E_j\right)\notag\\
	\overset{(b)}{=}&(1+o(1))\sum\limits_{i=1}^{k}\binom{n/d}{i} (-1)^{i+1}\binom{n-id}{s-id}\bigg/\binom{n}{s}\notag\\
	\overset{(c)}{\leq}&(1+o(1))\sum\limits_{i\text{ is odd}}^{i\leq k}\binom{n/d}{i}\frac{144s(n-id)}{(12s-1)(12(n-id)-1)} \left(\frac{n-id}{n}\right)^{n-id+0.5}\left(\frac{s}{s-id}\right)^{s-id+0.5}\left(\frac{s}{n}\right)^{id}-\notag\\
	&(1+o(1))\sum\limits_{i\text{ is even}}^{i\leq k}\binom{n/d}{i} \left(\frac{n-id}{n}\right)^{n-id+0.5}\left(\frac{s}{s-id}\right)^{s-id+0.5}\left(\frac{s}{n}\right)^{id}\notag\\
	\overset{(d)}{=}&(1+o(1))\sum\limits_{i\text{ is odd}}^{i\leq k}\binom{n/d}{i} \left(1+\beta'(n,k)\right)\left(1-\alpha'(n,k)\right)\left(\frac{s}{n}\right)^{kd}-\sum\limits_{i\text{ is even}}^{i\leq k}\binom{t}{k} \left(1-\alpha'(n,k)\right)\left(\frac{s}{n}\right)^{kd}\notag\\
	\overset{(e)}{=}&1-(1-\delta^d)^{n/d}-\sum\limits_{i=k+1}^{n/d} \binom{n/d}{i}(-1)^{i+1}\delta^{id}+o(1)\notag\\
	\overset{(f)}{=}&1-(1-\delta^d)^{n/d}+o(1)\notag\\
	\overset{(g)}{=}&o(1).
	\end{align}
	The above, step (a) is based on the symmetry of events $E_i$. Step (b) utilizes the definition of event $E_i$ and the structure of coding matrix $A^{\text{FRC}}$. Step (c) utilizes Sterlin's inequality (\ref{eq:sterlin}). In the step (d), since $k=n^{0.6}$, we have $id=o(n)$ for $1\leq i\leq k$ and
	\begin{equation*}
	\left(\frac{n-id}{n}\right)^{n-id+0.5}\left(\frac{s}{s-id}\right)^{s-id+0.5}=1-\alpha'(n,k) \text{ and } \lim\limits_{n\rightarrow \infty}\alpha'(n,k) = 0.
	\end{equation*}
	\begin{equation*}
	\frac{144s(n-id)}{(12s-1)(12(n-id)-1)}=1+\beta'(n,k) \text{ and } \lim\limits_{n\rightarrow \infty}\beta'(n,k) = 0.
	\end{equation*}
	Step (e) is based on the similar argument in the proof of (\ref{eq:same_pf_1}). The last step utilizes the fact that, when $i\geq n^{0.6}$ and $d=\log(n\log(1/\delta))/\log(1/\delta)$,
	\begin{equation*}
	\binom{n/d}{i}\delta^{id}\leq \left(\frac{en}{di}\delta^d\right)^i\leq  n^{-0.6n^{0.6}}.
	\end{equation*}
	The last step (g) is based on the choice of $d$ such that $(1-\delta^d)^{n/d}=e^{-1/\log(n\log(1/\delta))}=1-o(1)$. Therefore, the theorem follows.
\end{proof}

\subsection{Proof of Theorem~\ref{thm:approx_lower_bound}}

\begin{proof}
	Define the indicator function
	\begin{equation}
	X_i = \left\{\begin{matrix}
	1, \|A_{S,i}\|_0 =0\\ 
	0, \|A_{S,i}\|_0 >0
	\end{matrix}\right.
	\end{equation}
	Then we can obtain
	\begin{align}
	\mathbb{P}[\text{err}(A_S)>c]\geq\mathbb{P}\left[\sum\limits_{i=1}^nX_i>c\right].
	\end{align}
	Based on the similar proof of Lemma~\ref{lm:opt_structure}, we have
	\begin{align}
	\min\limits_{A\in\mathcal{A}_n^d}\mathbb{P}[\text{err}(A_S)>c]\geq\min\limits_{A\in\mathcal{U}_n^d}\mathbb{P}\left[\sum\limits_{i=1}^nX_i>c\right].
	\end{align}
	Suppose that
	\begin{equation}
	A^*=\argmin\limits_{A\in\mathcal{U}_n^d}\mathbb{P}\left[\sum\limits_{i=1}^nX_i>c\right].
	\end{equation}
	Based on the results in the Lemma~\ref{lm:independent_set}, we can construct a set $I_d^*$ such that $|\mathcal{I}_d^*|\geq \lfloor n/d^2\rfloor$ and 
	\begin{equation}
	\text{supp}(A_i^*)\cap \text{supp}(A_j^*)=\emptyset, \forall i\neq j, i,j\in\mathcal{I}_d.
	\end{equation}
	Therefore, we have
	\begin{align}
	\min\limits_{A\in\mathcal{A}_n^d}\mathbb{P}[\text{err}(A_S)>c]\geq\mathbb{P}\left[Y(I_d^*)>c\right],
	\end{align}
	where random variable $Y(I_d^*)=\sum\nolimits_{i\in I_d^*}X_i$. 
	
	Suppose that $d=o(s)$. Each indicator function $X_i$ is a Bernoulli random variable with
	\begin{equation}
	\mathbb{P}(X_i=1)=\binom{n-d}{s-d}\bigg/\binom{n}{s}\overset{(a)}{=}(1+o(1))\left(1-\frac{d}{n}\right)^{n-d+0.5}\left(1+\frac{d}{s-d}\right)^{s-d+0.5}\delta^d\overset{(a)}{=}(1+o(1))\delta^d.
	\end{equation}
	The above, step (a) utilizes sterlin's approximation; step (b) is based on the fact that $s=\Omega(1)$ and $d=o(s)$. First, the expectation of $Y(I_d^*)$ is given by
	\begin{equation}
	\mathbb{E}[Y(I_d^*)]=t(1+o(1))\delta^d.
	\end{equation}
	Furthermore, considering the fact that, for any $i,j\in I_d^*$ with $i\neq j$, the random variable $X_iX_j$ is a also Bernoulli random variable with
	\begin{equation}
	\mathbb{P}(X_iX_j=1)=\binom{n-2d}{s-2d}\bigg/\binom{n}{s}=(1+o(1))\left(1-\frac{2d}{n}\right)^{n-2d+0.5}\left(1+\frac{2d}{s-2d}\right)^{s-2d+0.5}\delta^{2d}=(1+o(1))\delta^{2d},
	\end{equation}
	the variance of $Y(I_d^*)$ is given by
	\begin{align}
	\text{Var}[Y(I_d^*)]=&\mathbb{E}[Y^2(I_d^*)]-\mathbb{E}^2[Y(I_d^*)]\notag\\
	=&\mathbb{E}\left[\sum\limits_{i\in I_d^*}X_i^2+2\sum\limits_{i,j\in I_d^*,i\neq j}X_iX_j\right]-\mathbb{E}^2[Y(I_d^*)]\notag\\
	=&(1+o(1))\left[t\delta^d+t(t-1)\delta^{2d}-t^2\delta^{2d}\right]\notag\\
	=&(1+o(1))\left[t\delta^d(1-\delta^d)\right]
	\end{align}
	Therefore, utilizing the Chebyshev inequality, we have the following upper bound.
	\begin{align}
	\mathbb{P}\left\{Y(I_d^*)\leq\mathbb{E}[Y(I_d^*)]-2\mathbb{E}^{0.5}[Y(I_d^*)]\right\}\leq&\mathbb{P}\left\{|Y(I_d^*)-\mathbb{E}[Y(I_d^*)]|\geq 2\mathbb{E}^{0.5}[Y(I_d^*)]\right\}\notag\\
	\leq&\frac{\text{Var}[Y(I_d^*)]}{4\mathbb{E}[Y(I_d^*)]}=(1+o(1))\frac{1-\delta^d}{4}.
	\end{align}
	Assume that $c\leq\mathbb{E}[Y(I_d^*)]-2\mathbb{E}^{0.5}[Y(I_d^*)]$, then we have
	\begin{align}
	\mathbb{P}\left\{Y(I_d^*)\leq c\right\}\leq&\mathbb{P}\left\{Y(I_d^*)\leq\mathbb{E}[Y(I_d^*)]-2\mathbb{E}^{0.5}[Y(I_d^*)]\right\}\leq 1/4.
	\end{align}
	This result implies that 
	\begin{equation}
	\mathbb{P}[\text{err}(A_S)>c]> \frac{3}{4},
	\end{equation}
	which is a contradiction. Therefore, the parameter $c$ should satisfy $c> \mathbb{E}[Y(I_d^*)]-2\mathbb{E}^{0.5}[Y(I_d^*)]$, which implies that 
	\begin{equation}
	\mathbb{E}[Y(I_d^*)]<2c+4.
	\end{equation} 
	Since $\mathbb{E}[Y(I_d^*)]=(1+o(1))t\delta^d$ and $\lfloor n/d^2\rfloor\delta^d$ is monotonically non-increasing with $d$, the minimum computation load should satisfy
	\begin{equation}
	d\geq\frac{\log(n\log^2(1/\delta)/(2c+4)\log^{2}(n/(2c+4)))}{\log(1/\delta)}.
	\end{equation}
	Therefore, the theorem follows.
\end{proof}

\subsection{Proof of Theorem~\ref{thm:raptor}}

We use the analysis of the decoding process as described
in~\cite{luby2001efficient}. Based on the choice of $b=\lceil 1/\log(1/\delta)\rceil+1$ and $\delta=s/n$, we can obtain that
\begin{equation}
\frac{n(1-2\epsilon)}{b(1-4\epsilon)}<n-s.
\end{equation}
Based on the results in~\cite{luby2001efficient}, to successfully recover $n/b(1-\epsilon)$ blocks from $\frac{n(1-2\epsilon)}{b(1-4\epsilon)}$ received results with probability $1-e^{-cn}$, we need to show the following inequality holds.
\begin{equation}
e^{-\frac{1-2\epsilon}{1-4\epsilon}\Omega'(x)}<1-x, \forall x\in [0,1-\epsilon],
\end{equation}
where $\Omega'(x)$ is the derivative of the generating function fo the degree distribution $P_w$. Note that
\begin{equation}
\Omega'(x)=\frac{1}{u +1}\left(u-\ln(1-x)+x^D-\sum\limits_{d=D+1}^{\infty}\frac{x^d}{d}\right)
\end{equation}
Utilizing the fact that $x^D>\sum\nolimits_{d=D+1}^{\infty} x^d/d$, the theorem follows.

\end{document}